\documentclass[aps,prx,preprint,bibliography,superscriptaddress]{revtex4-1}
\usepackage[utf8]{inputenc}
\usepackage{amsmath}
\usepackage{physics}
\usepackage{mathtools}
\usepackage{amsfonts}
\usepackage{amssymb}
\usepackage{graphicx}
\usepackage{subfigure}
\usepackage{textcomp}
\usepackage{color}
\usepackage[dvipsnames]{xcolor}
\usepackage{mathrsfs}

\begin{document}

\title{Nonlinear Hall effect in Rashba systems with hexagonal warping}
\author{Soumadeep Saha}
\affiliation{Undergraduate Programme, Indian Institute of Science, Bangalore 560012, India}
\author{Awadhesh Narayan}
\email{awadhesh@iisc.ac.in}
\affiliation{Solid State and Structural Chemistry Unit, Indian Institute of Science, Bangalore 560012, India}

\date{\today}

\begin{abstract}
Rashba spin-orbit coupled systems are an important class of materials noted for diverse fundamental and applied phenomena. Recently, the emergence of non-linear Hall effect under conditions of time-reversal symmetry has been discovered in materials with broken inversion symmetry. In this work, we study the second- and third-order Hall response in Rashba systems with hexagonal warping. Starting with a low-energy model, we obtain the analytic expressions and discover the unique dipole profile in Rashba systems with hexagonal warping. Furthermore, we extend the analysis using a realistic tight-binding model. Next, we predict the existence of a third-order Hall effect in these systems, and calculate the Berry connection polarizability tensor analytically. We also show how the model parameters affect the third-order conductivity. Our predictions can help in the experimental realization of Berry curvature multipole physics in Rashba materials with hexagonal warping, and provide a new platform for engineering the non-linear Hall effects.
\end{abstract}

\maketitle

\section{INTRODUCTION}

The celebrated Hall effect is a consequence of breaking of the time-reversal symmetry (TRS) in the linear response regime~\cite{cage2012quantum,thouless1982quantized}. Magnetic fields or magnetic dopants break TRS in materials leading to the appearance of the Hall effect~\cite{luttinger1958theory,chien2013hall}. Berry curvature, which is derived from Berry flux, plays the role of a magnetic field in momentum space, and can be understood to lead to Hall responses~\cite{vanderbilt2018berry}. Recently, pioneering work of Sodemann and Fu explored the role of the dipole moment of the Berry curvature, called the Berry curvature dipole (BCD) in transport properties of quantum systems~\cite{sodemann2015quantum}. BCD leads to non-linear Hall effects even in the presence of TRS, provided that the inversion symmetry of the crystal is broken~\cite{du2021nonlinear,ortix2021nonlinear}. The order of moment of the Berry curvature can be connected to the order of the Hall effect. The zeroth order moment is connected to the conventional Hall effect, whereas the second harmonics in Hall signals are caused by the first-order moment of Berry curvature, i.e., the BCD. Continuing further, higher order moments of Berry curvature are proposed to be connected to the appearance of higher harmonics of Hall signal. As an instance, the third-order Hall effect is described by another intrinsic geometric property of bands known as Berry connection polarizability (BCP) tensor~\cite{liu2022berry}. 

Non-linear Hall effect and its connection to BCD have been predicted for a variety of materials. These include, among others, two-dimensional materials such as transition metal dichalcogenides (TMDs), tilted massive Dirac and Weyl cone systems~\cite{du2018band,xiao2010berry,singh2020engineering,joseph2021topological}, strained monolayer and bilayer graphene~\cite{battilomo2019berry}, Weyl semimetals~\cite{yan2017topological}, buckled honeycomb lattices of silicene, germanene and stanene~\cite{bandyopadhyay2022electrically}, strong Rashba systems~\cite{facio2018strongly}, and Janus TMDs~\cite{joseph2021tunable}. Complementary developments have also been achieved in the experimental direction. The first observation of second-order Hall effect was made in few layer TMD, WTe$_2$~\cite{kang2019nonlinear, ma2019observation}. Following this discovery, non-linear Hall signals have also been reported in materials such as Dirac semimetals~\cite{shvetsov2019nonlinear}, Kondo materials~\cite{dzsaber2021giant}, WSe$_2$~\cite{qin2021strain,huang2020giant}, topological insulator surfaces~\cite{lv2021experimental}, and oxide interfaces of LaAlO$_3$/SrTiO$_3$~\cite{lebedev2021gate,lesne2022designing}, among others. Very recently, third-order Hall signals have been theoretically predicted in multi-weyl semimetals~\cite{roy2022non}, and detected experimentally in MoTe$_2$~\cite{lai2021third}.

An important group of materials for exploring fundamental spin-orbit physics and spintronic applications are materials with Rashba spin-orbit coupling~\cite{manchon2015new}. Recently, pressure driven phase transition have been discovered in three-dimensional bismuth tellurium iodine~\cite{bahramy2012emergence,facio2018strongly}, a material with a strong Rashba effect. Here the usual quadratic Hamiltonian is augmented by a linear-in-momentum term arising from the Rashba interaction. On the other hand, surface states of topological insulators such as Bi$_2$Te$_3$, Bi$_2$Se$_3$, and Sb$_2$Te$_3$ are described by a massless Dirac equation in the low energy regime~\cite{zhang2009topological}. These materials belong to a larger class of systems having single Dirac cone. Angle-resolved photoemission spectroscopy has revealed that the Fermi surface of these systems have a snowflake structure~\cite{fu2009hexagonal}. This uncommon structure is explained by the addition of a hexagonal warping term in the Hamiltonian, which manifests as a cubic-in-momentum term. These exciting developments suggest the need of exploring non-linear Hall response theoretically in a generalized framework, governed by a common Hamiltonian incorporating these different terms.

In this work, we study the second- and third-order Hall response in Rashba systems with hexagonal warping using a low energy model, as well as a tight-binding model. Through analytical calculations, we obtain the expressions for Berry curvature, and numerically calculate the BCD profile. We show that the unusual BCD profile obtained can be explained as an interplay of various terms in the Hamiltonian and the band structures arising from it. We also perform tight-binding calculations and find complementary results. We find that the particle-hole asymmetry plays an important role in the formation of an unusual BCD profile in these systems. Further, we calculate the BCP tensor components for the low energy model and predict the existence of a third-order Hall effect. We also study the effect of warping strength on the BCP tensor components. Furthermore, we explore third-order signals after introducing a band gap in the system, and find interesting connections between the warping strength and the gap parameter. Our work can help in experimental studies of Rashba systems with hexagonal warping, and potentially enable characterization of new materials belonging to this class.

The rest of the paper is organized as follows. In Section II, we give an overview of the relevant theory on Berry curvature, BCD, BCP and Rashba physics. Next we describe the results from the low-energy model in Section IIIA. In Section IIIB, we outline the tight-binding model and related calculations. Section IV discusses the third-order Hall response and the conductivity tensor. Finally, Section V provides experimental consideration of the effects investigated in the manuscript.

\section{BERRY CURVATURE, BERRY CURVATURE DIPOLE, BERRY CONNECTION POLARIZABILITY TENSOR AND RASHBA EFFECT}

In this section, we briefly review the central theoretical concepts of Berry curvature, BCD and BCP tensor, along with their role in transport properties. When an oscillating electric field, $\mathbf{E}(t)=\textrm{Re}[\epsilon e^{i\omega t}]$ is applied, a non-linear current flows through the material, given by $J_a=\textrm{Re}[J_a^{(0)}+J_a^{(2)}e^{2i\omega t}]$, with $J_a^{(0)}=\chi_{abc}^{(0)}\epsilon_b\epsilon_c^*$ and $J_a^{(2)}=\chi_{abc}^{(2)}\epsilon_b\epsilon_c$. The coefficients $\chi_{abc}^{(0)}$ and $\chi_{abc}^{(2)}$ are found to be equal for systems with time-reversal symmetry, and are given by~\cite{sodemann2015quantum}

\begin{equation}
    \chi_{abc}^{(0)}=\chi_{abc}^{(2)}=\frac{\epsilon^{acd}D_{bd}e^3\tau}{2\hbar^2(1+i\omega t)},
\end{equation}

where $\epsilon^{acd}$ is the Levi-Civita antisymmetric tensor, $\omega$ is the frequency, $-e$ is the charge of the electron, and $\tau$ is the relaxation time. $D_{bd}$ is the BCD, which is given by 

\begin{equation}
    D_{bd}=-\sum_i\int[d\mathbf{k}]\frac{\partial \epsilon^i_\mathbf{k}}{\partial k_b}\Omega^d_{i\mathbf{k}}\frac{\partial f_\mathbf{k}}{\partial\epsilon^i_\mathbf{k}}.
    \label{Eqn:BCD}
\end{equation}

Here $f_\mathbf{k}$ denotes the equilibrium Fermi-Dirac distribution and $a,b,c,d\in \{x,y,z\}$. Further, $\epsilon^i_\mathbf{k}$ is the energy of the $i^{th}$ band with wavevector $\mathbf{k}$, and $[d\mathbf{k}]=d^3\mathbf{k}/(2\pi)^3$. The kernel of the integral of Equation~\ref{Eqn:BCD} is the BCD density~\cite{du2018band},

\begin{equation}
   d_{bd}(\mathbf{k})=(\partial ^b_\mathbf{k}\epsilon_\mathbf{k})\Omega^d_\mathbf{k}.
    \label{Eqn:BCDD}
\end{equation}

The fundamental quantity of interest, appearing in Equation~\ref{Eqn:BCD} and Equation~\ref{Eqn:BCDD} is the Berry curvature, $\Omega^d_{i\mathbf{k}}$, which stands for the Berry curvature component in the direction $d$ for the $i$-th energy band with wavevector $\mathbf{k}$. Berry curvature, in turn, can be found as~\cite{xiao2010berry}

\begin{equation}
    \Omega^a_{i\mathbf{k}}=-\epsilon^{abc}\textrm{Im}\sum_{i\neq j}\frac{\langle i|\frac{\partial H}{\partial k_b}|j\rangle\langle j|\frac{\partial H}{\partial k_c}|i\rangle-\langle i|\frac{\partial H}{\partial k_c}|j\rangle\langle j|\frac{\partial H}{\partial k_b}|i\rangle}{(\epsilon^i_\mathbf{k}-\epsilon^j_\mathbf{k})^2},
    \label{BC}
\end{equation}

where $|i\rangle$ is eigenstate of the Hamiltonian $H$, corresponding to the band $i$, having energy $\epsilon^i_\mathbf{k}$ and wavevector $\mathbf{k}$. The integral in Equation~\ref{Eqn:BCD} survives even in the case of time-reversal symmetric systems, provided inversion symmetry is broken. This breaking of symmetry allows for a non-zero value of BCD for cases with finite Berry curvature.

The third-order Hall effect can be connected to the BCP tensor, an intrinsic geometric property of bands. The BCP tensor components are given by~\cite{lai2021third}

\begin{equation}
    G_{ab}=2 \textrm Re\sum_{i\neq j}\frac{(\mathcal{A}_a)_{ij}(\mathcal{A}_b)_{ji}}{\epsilon_i-\epsilon_j},
    \label{Eqn:BCP}
\end{equation}

with $\mathcal{A}$ being the Berry connection. When an electric field is applied, the Berry curvature gets perturbed, and needs to be corrected to first order as~\cite{liu2022berry}

\begin{equation}
    \Tilde{\Omega}(\mathbf{k})=\Omega(\mathbf{k})+\Omega^{(1)}(\mathbf{k}).
\end{equation}

The first order correction is given by 

\begin{equation}
    \Omega^{(1)}=\nabla_\mathbf{k}\times \mathcal{A}^{(1)},
\end{equation}

and the first order correction of the Berry connection, $\mathcal{A}^{(1)}$ can be related to the BCP tensor through

\begin{equation}
    \mathcal{A}^{(1)}_a (\mathbf{k})=G_{ab}(\mathbf{k})E_b.
\end{equation}

Berry connection polarizability is then defined as

\begin{equation}
    P_{ab}=\frac{\partial\Tilde{\Omega}_a}{\partial E_b}=\epsilon_{acd}\partial_c G_{db}.
\end{equation}

The BCP components can be used to evaluate the third-order conductivity tensor $\chi$. Continuing from the second-order response, the third-order response is related to the electric field components as $j_{a}^{(3)}=\chi_{abcd}E_b E_c E_d$, assuming Einstein summation convention. In the case of third-order response, $\chi$ can be separated into two parts, $\chi^I$ being linear in $\tau$, and $\chi^{II}$ being cubic in $\tau$. The conductivity components are given by

\begin{subequations}
\begin{equation}
    \chi_{abcd}^I=\tau \left(2\int \left[ d\mathbf{k}\right](\partial_a \partial_b G_{cd})f_\mathbf{k}-\int \left[ d\mathbf{k}\right](\partial_c \partial_d G_{ab})f_\mathbf{k}-\frac{1}{2}\int \left[ d\mathbf{k}\right](v_a v_b G_{cd})f^{''}_\mathbf{k} \right),
\end{equation}
\begin{equation}
     \chi_{abcd}^{II}=-\tau^3 \int\left[ d\mathbf{k}\right]v_a\partial_b\partial_c\partial_d f_\mathbf{k},
\end{equation}
\end{subequations}
where $v_i=\frac{\partial E}{\partial k_i}$ is the group velocity. We will use this formalism to investigate the non-linear Hall properties in our Rashba systems with hexagonal warping.

\section{RESULTS AND DISCUSSION}

\subsection{Low-energy model}

First, let us present a low-energy model to describe our Rashba system with hexagonal warping. For the rest of the discussion, we set $\hbar=e=1$. The Hamiltonian reads~\cite{lesne2022designing}

\begin{equation}
\mathscr{H}=\frac{\textbf{k}^2}{2m}-\alpha_R(\sigma_xk_y - \sigma_yk_x)+\frac{\lambda}{2}(k^3_+ + k^3_-)\sigma_z,
\end{equation}

where $\alpha_R$ is the Rashba coefficient, $\lambda$ is the strength of warping, $m$ is the effective mass of electron, and $k_{\pm}=k_x\pm ik_y$. Let us recall the symmetries of this Hamiltonian, which will also be important for subsequent discussions. Under threefold rotation operator $C_3$ and mirror symmetry operator $M_x$, the momentum $k$ and Pauli spin matrices $\sigma_x, \sigma_y$ and $\sigma_z$ transform as~\cite{lesne2022designing,zhou2020highly,fu2009hexagonal}

\begin{equation}
   \begin{aligned}
    C_3: k_{\pm}\rightarrow e^{\pm2\pi i/3}k_{\pm},\quad\sigma_{\pm}\rightarrow e^{\pm2\pi i/3}\sigma_{\pm},\quad \sigma_z\rightarrow\sigma_z,
    \\
    M_x: k_+\rightarrow-k_-,\quad\sigma_x\rightarrow\sigma_x,\quad\sigma_{y,z}\rightarrow-\sigma_{y,z}.
    \end{aligned}
\end{equation}

Along with the spatial symmetry constraints, there are constraints from time-reversal symmetry. The time-reversal symmetry operator is represented as $\mathcal{T}=i\sigma_y\mathcal{K}$, where $\mathcal{K}$ is the complex conjugation operator. The Hamiltonian should be invariant under time reversal operator, which translates to the constraint $\mathcal{H}(\mathbf{k})=\mathcal{T}\mathcal{H}(-\mathbf{k})\mathcal{T}^{-1}=\sigma_y\mathcal{H}^*(-\mathbf{k})\sigma_y$.

At the leading order, the breaking of threefold rotation symmetry for the system is modeled by assuming unequal coefficients for spin-orbit coupling terms which are linear in momentum. The coefficients of $k_x$ and $k_y$ are then taken to be $v_x$ and $v_y$, respectively. Thus, our modified Hamiltonian now becomes

\begin{equation}
\mathscr{H}=\frac{k^2_x + k^2_y + k^2_z}{2m}-(v_y\sigma_xk_y - v_x\sigma_yk_x)+\frac{\lambda}{2}(k^3_+ + k^3_-)\sigma_z.
\label{Eqn:Low_Energy_Ham}
\end{equation}

The corresponding dispersion relation is given by

\begin{equation}
E_{\pm}(\textbf{k})=\frac{k^2_x + k^2_y + k^2_z}{2m}\pm\sqrt{v^2_xk_x^2+v^2_yk_y^2+k^6_x\lambda^2-6k_x^4k_y^2\lambda^2+9k_x^2k_y^4\lambda^2}.
\end{equation}

We note that the bands touch each other only at $k_x=k_y=k_z=0$ in the momentum space. Energy is measured in units of $k_F^2$, and hence value of $m$ is chosen to be 0.5. For this dispersion relation, $k_z$ only provides a constant shift in the energy value, and does not affect the band structure for the system. Thus, later we will choose to set $k_z=0$, which allows us to describe our model as an effective two-dimensional one. 

We begin our analysis by deriving analytical expressions for Berry curvature. The Berry curvature components along the three cartesian directions read

\begin{subequations}
    \begin{equation}
        \Omega^x_\pm=0,
    \end{equation}
    \begin{equation}
        \Omega^y_\pm=0,
    \end{equation}
    \begin{equation}
        \Omega^z_\pm=\frac{\pm v_xv_yk_x(-3k_y^2+k_x^2)\lambda}{(v_x^2k_x^2+v_y^2k_y^2+k_x^2(k_x^2-3k_y^2)^2\lambda^2)^{3/2}}.
    \end{equation}
\end{subequations}

We note that $\Omega^z_\pm$ is independent of $k_z$. This dependence of $\Omega^z_\pm$ on only $k_x$ and $k_y$ motivates us to write $\Omega^z_\pm$ in polar form as

\begin{equation}
\Omega^z_\pm(k,\theta)=\pm\frac{2\sqrt{2}\lambda v_xv_yk^3\cos{3\theta}}{[2k^2(v_x^2\cos^2{\theta}+v_y^2\sin^2{\theta})+\lambda^2k^6\cos{6\theta}+\lambda^2k^6]^{3/2}},
\end{equation}

where $k=\sqrt{k_x^2+k_y^2}$, and $\theta$ is taken with reference from the $k_x$ direction.

Symmetry properties possessed by $\Omega^z$ are: (a) $\Omega^z$ is odd in $k_x$ , i.e., $\Omega^z(-k_x, k_y)=-\Omega^z(k_x, k_y)$. (b) $\Omega^z$ is even in $k_y$ ,i.e., $\Omega^z(k_x, -k_y)=\Omega^z(k_x, k_y)$. (c) It vanishes along the high symmetry directions of $k_x =0, k_x =\pm\sqrt{3}k_y$. These symmetry directions correspond to the directions where the cubic term in the Hamiltonian, namely $\frac{\lambda}{2}(k_+^3+k_-^3)$ vanishes. As noted, this term is the hexagonal warping term, and is essential for occurrence of Berry curvature in the systems under consideration. The effect of this term is thus directly manifested in the expression for Berry curvature, and provides a good connection between underlying physics and mathematical symmetry. \\

\begin{figure}[t] 
\includegraphics[width=\textwidth]{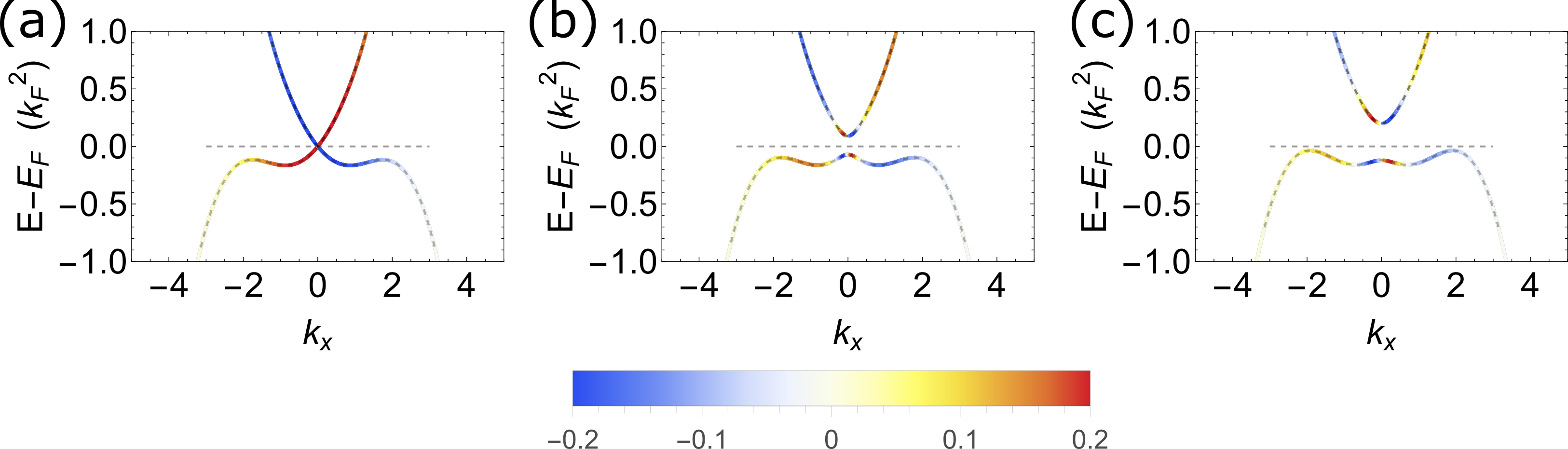}
    \caption{ \textbf{Band structure and Berry curvature for low energy Hamiltonian.} The band structure superimposed with the Berry curvature along $k_x$ (setting $k_z=0$), for (a) $k_y=0$, (b) $k_y=0.2$, and (c) $k_y=0.4$. The warping effect of band structure can be clearly observed from the plots, predominant in the valence band. There is a slight asymmetry in the distribution of Berry curvature along $k_x$, prominently seen in (c). We observe that Berry curvature changes rapidly near the band edges, which is crucial for the dipole profile.  Other parameters are chosen to be $\lambda=0.1$, $v_y=v_x=0.4$, and $m=1/2$.}
    \label{fig:Low_energy_Bands} 
\end{figure}

The band structures along $k_x$ (for $k_z=0$), are plotted in Fig.~\ref{fig:Low_energy_Bands} for varying $k_y$. The value of $z$ component of Berry curvature is superimposed on the band structure. As noted earlier, the bands are degenerate at the origin, and rapid changes in Berry curvature can be observed around $k_x=0$. The band structure is not symmetric about the zero energy, because of the particle-hole asymmetry, which is reflected by the $\frac{k^2}{2m}$ term in the Hamiltonian. A slight asymmetry of Berry curvature distribution on the band structure can also be observed along $k_x$, especially as $k_y$ becomes non-zero. This is a combined effect of the Rashba coefficients as well as the warping term $\lambda$. This deviation is necessary for the occurrence of the dipole profile, as we will discuss later.

\begin{figure}[t]
    \includegraphics[width=0.8\textwidth]{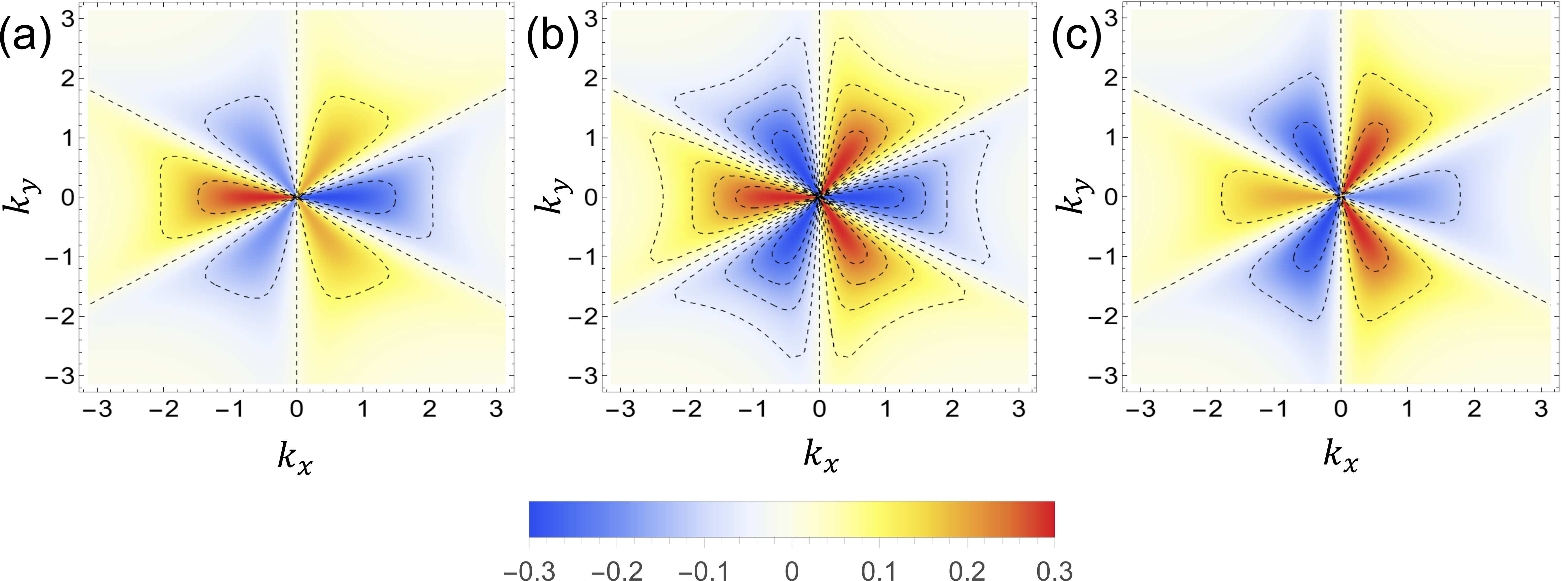}
    \caption{ \textbf{Distribution of Berry curvature.} Density plots of Berry curvature with hexagonal warping for (a) $v_y/v_x = 1.2$, (b) $v_y/v_x = 1$, and (c) $v_x/v_y = 1.2$. The effect of warping is clear from the plots, where the Berry curvature is mostly concentrated in six regions. Berry curvature is zero along the dashed straight lines, as expected from the symmetries of the Hamiltonian. The relative concentration of Berry curvature in different regions is determined by the ratio of Rashba coefficients. Other parameters common to the plots are $m=0.5$ and $\lambda=0.1$.}
    \label{fig:BC_contour}
\end{figure}

The distribution of the $z$ component of Berry curvature in $k_x-k_y$ plane is shown in Fig.~\ref{fig:BC_contour}. Except at the origin, where the bands are degenerate, Berry curvature is well-defined in the $k_x-k_y$ plane. Hexagonal warping forces Berry curvature to get distributed into six regions, as can be observed from the plots. Berry curvature vanishes along the mirror lines, which is expected from symmetry considerations. However, the relative strengths of Berry curvature in each region is determined by the ratio of Rashba coefficients $v_x$ and $v_y$. Thus, there are two independent, but essential factors which determine the distribution of Berry curvature for our system.

The rapidly changing Berry curvature near the band edges and the asymmetry of the band structure suggest the presence of BCD for the material systems described by our Hamiltonian. In these systems, because of the Rashba spin-orbit coupling, there is only one mirror symmetry, $M_x$. The presence of $M_x$ symmetry leads to $k_x$ and $k_y$ transforming as odd and even parameters, respectively. Thus, considering two-dimensional system $(k_z=0)$, only non-zero component of Berry curvature is the $z$ component, $\Omega^z$. Noting that $\Omega^z$ is odd in momentum, we find that $D_{xz}$ is the only non-zero component of BCD, owing to the symmetry considerations.

We address the second order nonlinear conductivity using Eq. 1. In typical materials, the relaxation time $\tau$ is of the order of picoseconds, and typically the frequency of alternating electric field is in the range 10-1000 Hz. In this regime, $\omega \tau <<1$ . Thus, in effect from Eq. 1, we get that second order nonlinear coefficients ($\chi$) are directly proportional to the Berry curvature dipole. Specifically, we have

\begin{equation}
   \chi_{xxy}\propto D_{xz}, \quad \chi_{yxx}\propto -D_{xz}.
\end{equation}

\begin{figure}[t] 
    \includegraphics[width=\textwidth]{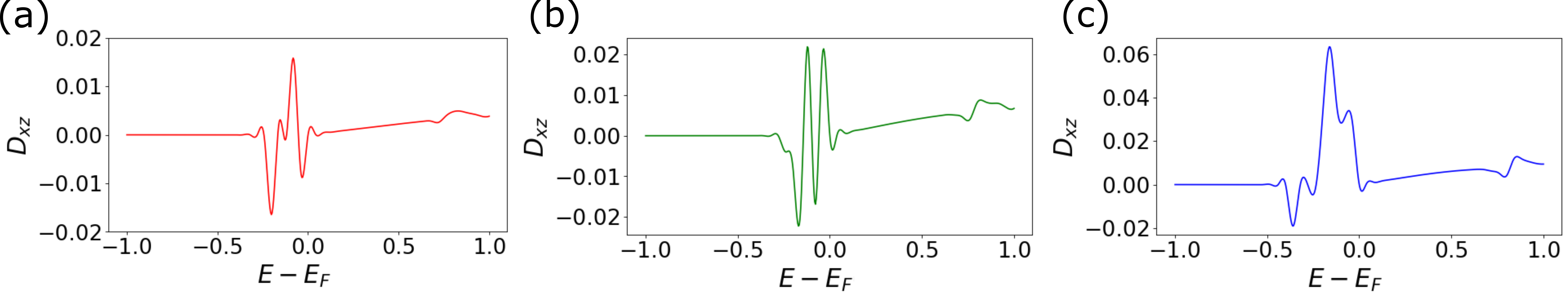}
    \caption{\textbf{Berry curvature dipole for the low energy model.} The non-zero component $D_{xz}$ as a function of energy, for varying warping strengths (a) $\lambda=0.1$, (b) $\lambda=0.2$, and (c) $\lambda=0.3$. $D_{xz}$ increases nearly linearly at positive values of $E-E_F$. For negative values of $E-E_F$, the value of $D_{xz}$ is vanishing away from the Fermi level. Near the Fermi level, there are multiple peaks and dips arising from rapid changes in the value of the Berry curvature. We notice the difference between maximum positive and negative values of BCD, caused by warping and the influence of the group velocity.}
    \label{fig:BCD}
\end{figure}

The quantitative details of conductivity depend on the properties of the material system under consideration. However, from qualitative point of view, the dipole profiles (Fig. \ref{fig:BCD}) provides a good picture of the behavior of $\chi$ as a function of Fermi energy.\\

Next, we calculate the BCD component $D_{xz}$ using Equation~\ref{Eqn:BCD}, as a function of energy. The plots are shown in Fig.~\ref{fig:BCD}, and present several interesting features that can be understood from the interplay of the band structure and the Berry curvature. We notice that for negative values of $E-E_F$, $D_{xz}$ is vanishing away from the Fermi level, starting from approximately $E-E_F=-0.25$. Observing the band structures (Fig.~\ref{fig:Low_energy_Bands}) for warping strength $\lambda=0.1$, we find that there are very small values of Berry curvature present at these energies. This explains the smooth vanishing of BCD for large negative values of energy.

For positive values of energy, there is a gradual linear increase in $D_{xz}$. This finite value of $D_{xz}$ arises because of the presence of substantial Berry curvature in the conduction band. The concentration of Berry curvature decreases away from the origin, but as pointed out before, there is a slight asymmetry in its distribution along $k_x$. This asymmetry is the cause of the linear increase in an otherwise stable value of BCD for positive values of $E-E_F$.

Near the Fermi level, there are multiple peaks and dips occurring for BCD, which can be understood from the rapid changes of the Berry curvature in this energy region. In addition to Berry curvature, the group velocity proportional to $\frac{\partial E}{\partial k_x}$ plays an important role in determining the dipole profile. At the band edges, the group velocity vanishes, forcing the dipole to vanish. This results in multiple sign changes in BCD at negative values of $E-E_F$ very close to the Fermi level. Near the band edges, group velocity changes sign rapidly, which can be qualitatively visualized through the change in the curvature of the band structure in this energy window. When integrated over the Brillouin zone, these small sign changes lead to large fluctuations in BCD values near the Fermi level, as seen in Fig.~\ref{fig:BCD}.

From Fig.~\ref{fig:BC_contour}, we note that the warping strength $\lambda$ plays a very sensitive role in determining the BCD. The reason for this is two-fold. Firstly, the distribution of Berry curvature caused by hexagonal warping (as shown in Fig.~\ref{fig:BC_contour}), and the warping of band structure itself, leads to rapid changes in the group velocity. Here we find a generic feature of $D_{xz}$ for these systems -- the BCD is neither symmetric, nor anti-symmetric with respect to $E-E_F$. This is an unusual behavior of BCD amongst the materials well-studied in literature~\cite{du2018band,zhang2018berry,xu2018electrically,battilomo2019berry}.

\begin{figure}[t]
    \centering
    \includegraphics[width=0.6\textwidth]{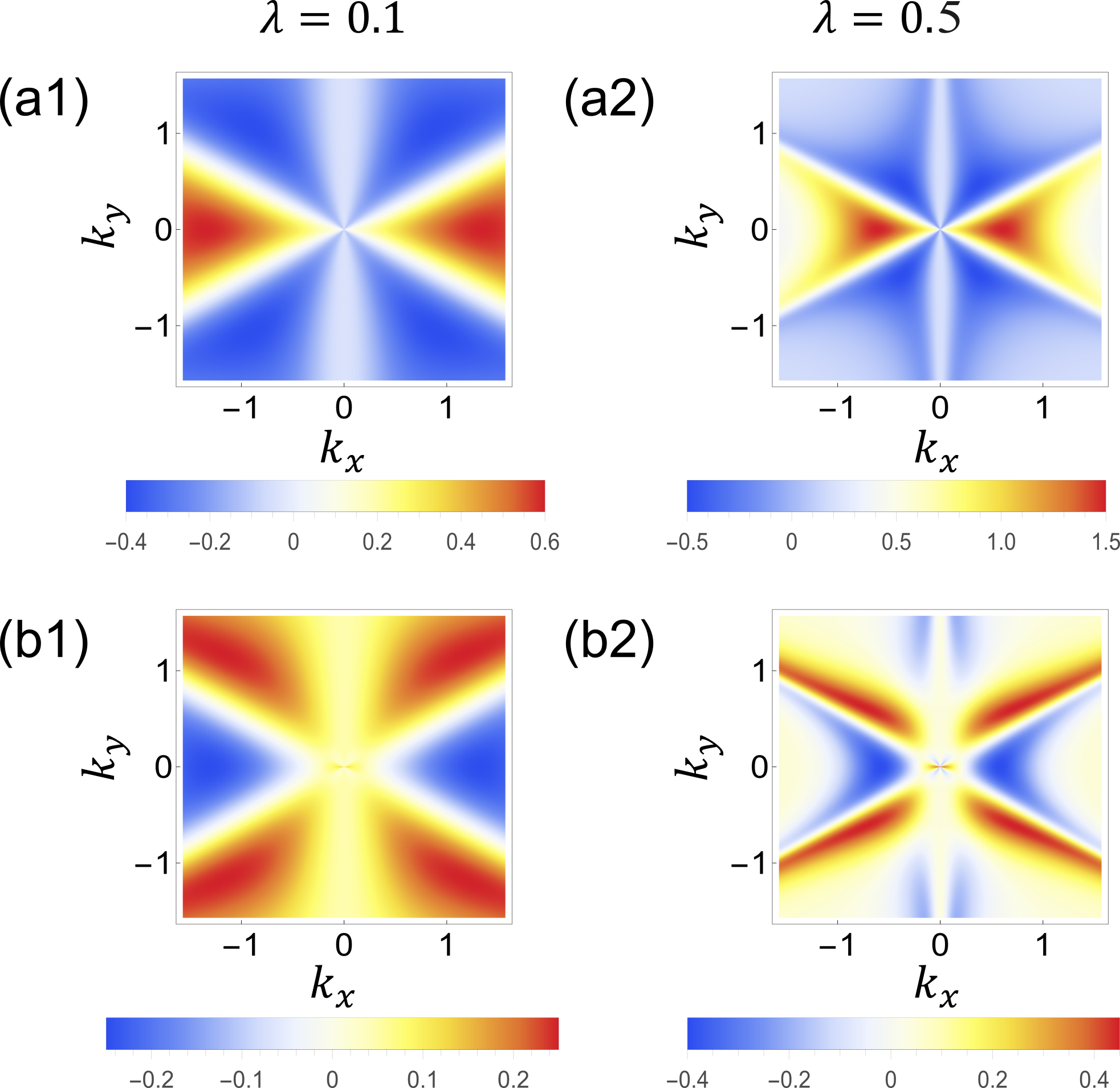}
    \caption{\textbf{Berry curvature dipole density.} BCD density plotted for the conduction band (panel (a)) and valence band (panel (b)), plotted in the $k_x-k_y$ plane ($k_z=0$), for different values of warping strength $\lambda$. The distribution of dipole density is symmetric along $k_x$ and $k_y$. An increase in warping causes the features to get more concentrated near the origin, because of increase in changes in group velocity. It is worth noting that the positive and negative values of density are not symmetric for conduction band, whereas they are almost symmetric for valence band. This distribution of the BCD density, along with group velocity effects results in a finite dipole. Here $m$ is taken to be 0.5.}
    \label{fig:BCDD}
\end{figure}

To gain further insights into BCD of hexagonally warped Rashba systems, we next look into the BCD density. Analytically, we find that $d_{xz}\sim \pm k_x^2(k_x^2-3k_y^2)$, thus being symmetric in both $k_x$ and $k_y$. The dipole density, $d_{xz}$, for conduction band and valence band is plotted in Fig.~\ref{fig:BCDD}. It is worth noting that as $\lambda$ increases, the density gets more concentrated near the origin. This results in a higher BCD occurring around low negative values of $E-E_F$, as shown in Fig.~\ref{fig:BCD}. However, since the density is distributed symmetrically in the $k_x-k_y$ plane, the unusual dipole profile is essentially caused by the warping of band structure itself, with the magnitude being largely dictated by the warping strength.

The density distribution is seen to be largely unaffected by the ratio of $v_x$ and $v_y$. However, the positive and negative values of BCD density are not equal in magnitude, which results in the net dipole being finite. To get a better understanding of the BCD, we employ next tight-binding models, which provide a more realistic description of material systems.

\subsection{Tight-binding model}

After finding several interesting behaviors of BCD and Berry curvature in the low-energy model for our system, we now employ a tight-binding model~\cite{lima2022tight,moca2005longitudinal,ast2012s,van2014tight} derived from the low-energy Hamiltonian. The tight-binding Hamiltonian reads

\begin{equation}
 \mathscr{H}=  \frac{1}{2m} (6-2\cos{k_x}-2\cos{k_y}-2\cos{k_z})-(v_y\sin{k_y}\sigma_x- v_x\sin{k_x}\sigma_y) + \lambda\sin{k_x}(6\cos{k_y}-2\cos{k_x}-4)\sigma_z,
\end{equation}

where $v_x$ and $v_y$ are the Fermi velocities in the $x$ and $y$ directions, respectively, $\lambda$ is the strength of warping, and $m$ is the effective mass of the electron.

 For numerical calculations, we use PythTB~\cite{yusufaly2013tight} package for the generation of the lattice system. The output from PythTB is then fed into the Wannier-Berri package~\cite{tsirkin2021high} for calculating band structures, Berry curvature and BCD. For ease of calculation, we set $v_x=v_y=1$, and the on-site energy to zero. In this setting, $\lambda$ becomes proportional to the hopping parameter $t$ ($t=1$), in units of which energy is measured. For calculation of BCD, the convergence was checked on a $100\times100\times100$ $k$-grid.

\begin{figure}[htp]
    \centering
    \includegraphics[width=0.8\textwidth]{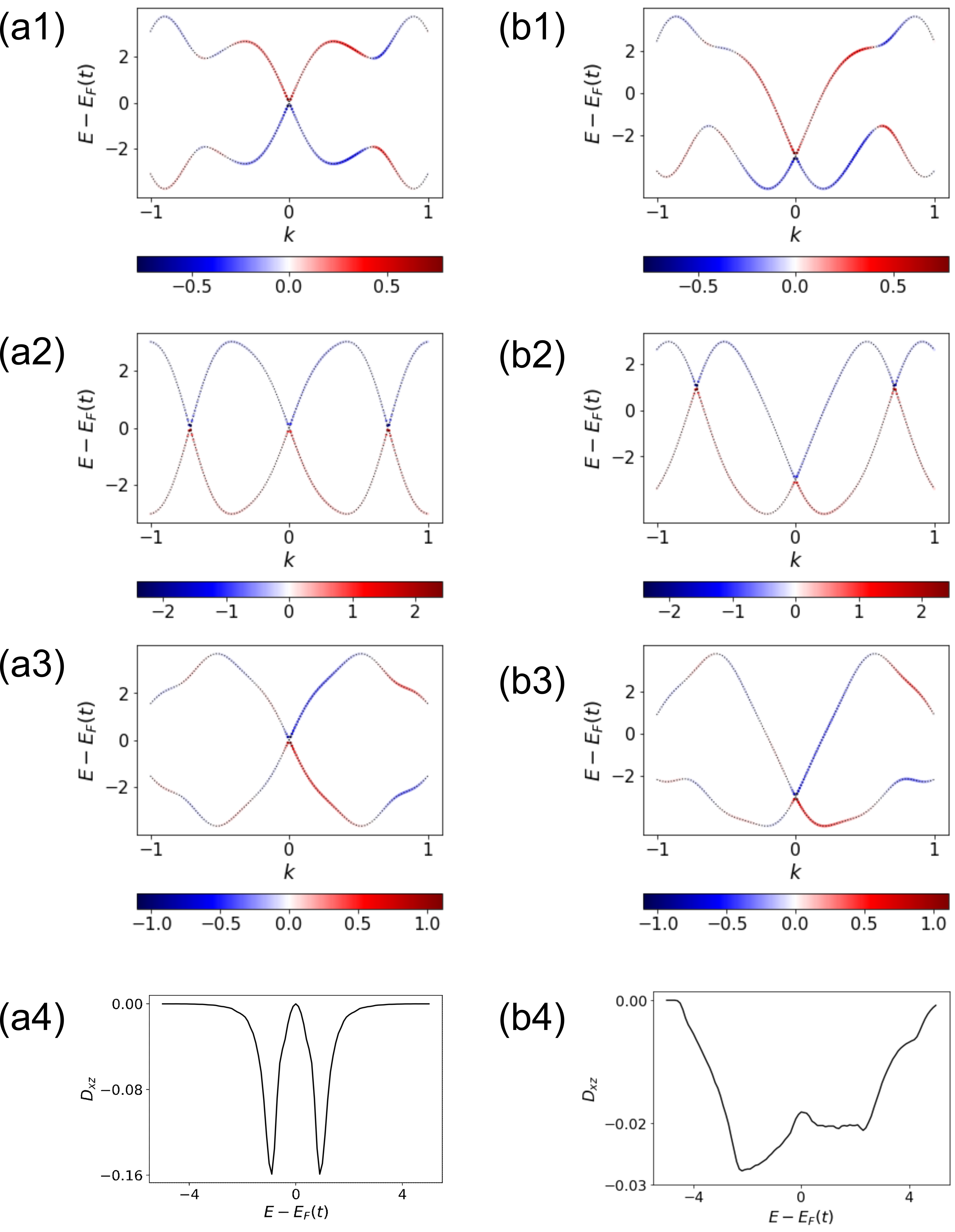}
    \caption{\textbf{Band structure, Berry curvature and Berry curvature dipole for tight-binding model.} Panel (a) corresponds to the case with particle-hole symmetry. The bands are plotted along the path $k_y=k_x (\tan \theta)$, with $\theta=30^{\circ}$ for (a1), $\theta=45^{\circ}$ for (a2), $\theta=60^{\circ}$ for (a3). Panel (b) corresponds to the case with particle-hole asymmetry. The bands are plotted along the path $k_y=k_x(\tan \theta)$, with $\theta=30^{\circ}$ for (b1), $\theta=45^{\circ}$ for (b2), $\theta=60^{\circ}$ for (b3). Figures (a4) and (b4) show the plot of dipole component, $D_{xz}$, as a function of the Fermi Energy. (a4) is symmetric about the energy axis as expected. Other parameters common to the plots are $\lambda=0.2$ and $v_y=v_x=1$.}
    \label{fig:TB}
\end{figure}

Our tight-binding model provides a good description of the effect of particle-hole asymmetry on the BCD of these systems. Fig.~\ref{fig:TB} shows the band structure with superimposed Berry curvature, along with the BCD for the tight-binding model, both in the presence and absence of the particle-hole asymmetry. The band structures show the Berry curvature being concentrated near the band edges, with maximum values for region around $k_x=k_y$, as can be observed from band structures in Fig.~\ref{fig:TB}(a2) and Fig.~\ref{fig:TB}(b2), consistent with the low-energy model. Fig.~\ref{fig:TB}(a4) shows the dipole profile when particle-hole symmetry is present. As expected, the profile is symmetric with respect to $E-E_F=0$. Moreover, $D_{xz}$ goes to zero when $E=E_F$. This is because of the exact cancellation of the dipole density at that energy.

It is worth noting that as particle-hole asymmetry is introduced, the band structures also become asymmetric with respect to Fermi energy, as observed in Fig.~\ref{fig:TB}(b1)-(b4). This affects the dipole profile, causing a larger magnitude of dipole for negative values of Fermi energies [see Fig.~\ref{fig:TB}(b4)]. We also note a near constant dipole value followed by a linear increase of BCD for positive values of $E-E_F$ in Fig.~\ref{fig:TB}(b4). This can be understood by the decrease of Berry curvature in the conduction band when asymmetry is present. Our analysis suggests that particle-hole asymmetry, a property often overlooked, plays a crucial role in determining the BCD in these systems, in addition to the important role of hexagonal warping.

\section{Third-order Hall response}

A natural extension of second-order non-linear Hall response is to next higher order, i.e., to investigate the third-order Hall response. So far, only a handful of systems have been identified to exhibit the third-order response. We next explore whether our Rashba systems with hexagonal warping can show such effects. Third-order Hall response is understood in the formalism of the BCP tensor~\cite{lai2021third,liu2022berry}. For our low-energy Hamiltonian (Equation~\ref{Eqn:Low_Energy_Ham}), using Equation~\ref{Eqn:BCP}, we venture to calculate the different BCP components for our Rashba system. We obtain them to be 

\begin{subequations}
\begin{equation}
G_{xx}=-\frac{4v_x^2k_x^6\lambda^2+v_y^2k_y^2(v_x^2+9(k_x^2-k_y^2)^2\lambda^2)}{4(v_x^2k_x^2+v_yk_y^2+k_x^2(k_x^2-3k_y^2)^2\lambda^2)^{5/2}},
\end{equation}

\begin{equation}
G_{yy}=-\frac{k_x^2(36v_x^2k_x^2k_y^2\lambda^2+v_y^2(v_x^2+(k_x^2+3k_y^2)^2\lambda^2))}{4(v_x^2k_x^2+v_yk_y^2+k_x^2(k_x^2-3k_y^2)^2\lambda^2)^{5/2}},
\end{equation}

\begin{equation}
G_{xy}=G_{yx}=\frac{12v_x^2k_x^5k_y\lambda^2+v_y^2k_xk_y(v_x^2+3(k_x^4+2k_x^2k_y^2-3k_y^4)\lambda^2)}{4(v_x^2k_x^2+v_yk_y^2+k_x^2(k_x^2-3k_y^2)^2\lambda^2)^{5/2}},
\end{equation}

\begin{equation}
G_{xz}=G_{yz}=G_{zx}=G_{zy}=G_{zz}=0.
\end{equation}
\end{subequations}

We find that out of the nine components of BCP tensor, only four components are non-zero. Further, the off-diagonal components are symmetric. From the expressions of BCP components, we notice that the denominator goes to zero as $k_x,k_y\rightarrow0$. This forces $G_{xx}$, $G_{yy}$ and $G_{xy}$ to diverge near the origin, hindering a complete analysis and understanding of the physical aspects of our hexagonally-warped Rashba systems.

\begin{figure}[t]
   \centering
    \includegraphics[width=\textwidth]{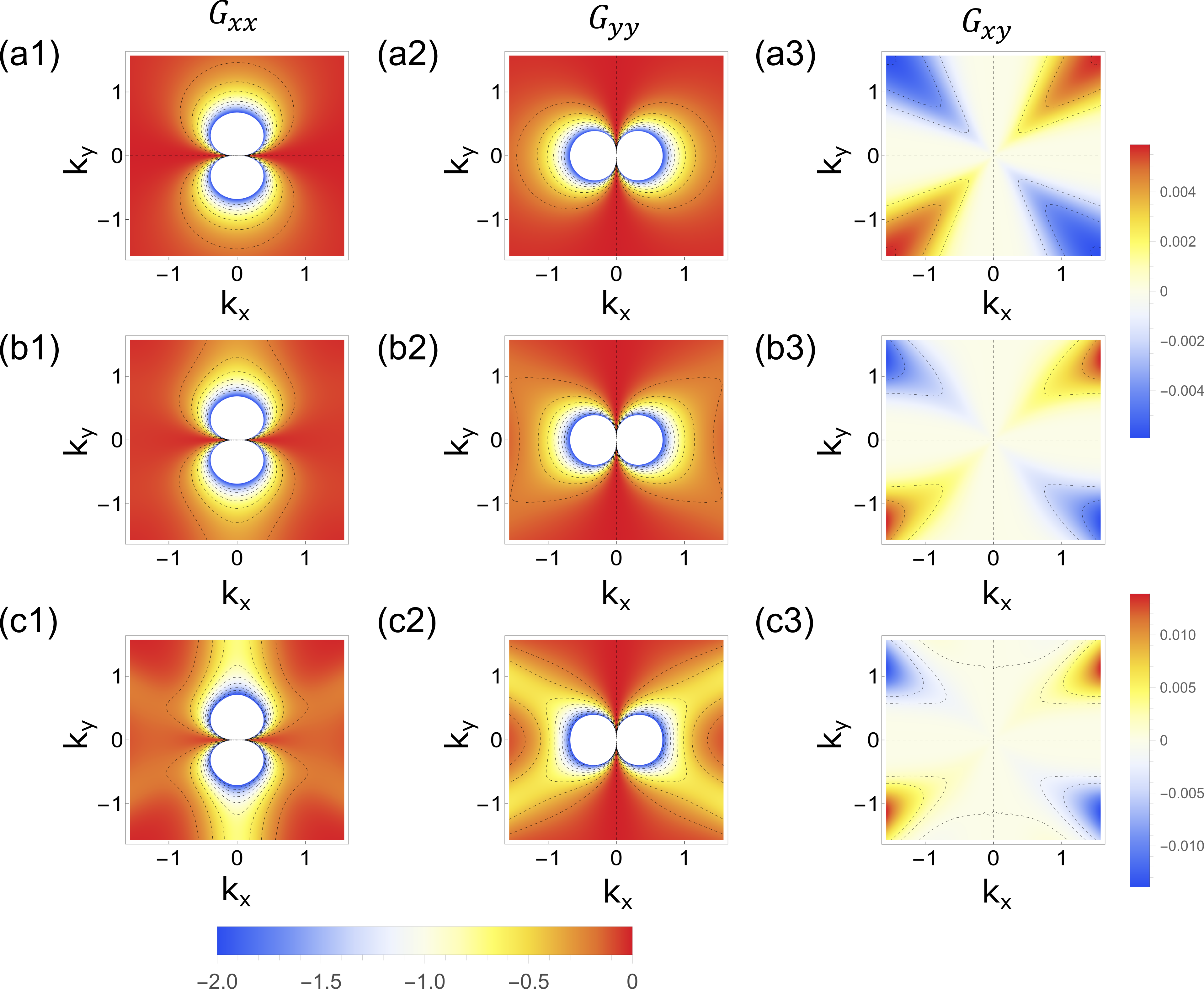}
    \caption{\textbf{Berry connection polarizability tensor components.} Distribution of Berry connection polarizability tensor components plotted in the $k_x-k_y$ plane (with $k_z=0$). $G_{xx}$ is plotted for (a1) $\lambda=0$, (b1) $\lambda=0.05$, (c1) $\lambda=0.1$; $G_{yy}$ for (a2) $\lambda=0$, (b2) $\lambda=0.05$, (c2) $\lambda=0.1$, and $G_{xy}$ for (a3) $\lambda=0$, (b3) $\lambda=0.05$, (c3) $\lambda=0.1$. $G_{xx}$ and $G_{yy}$ have a double-lobed structure, with `legs' forming away from origin as warping increases. These `legs' have the same sign of the BCP component. $G{xy}$ also forms `legs', but these have different signs. The distribution of BCP components near the origin is not much affected by warping, as can be noticed from all the plots.}
   \label{fig:Non_gapped_BPT}
 \end{figure}

$G_{xx}$ and $G_{yy}$ both are even in $k_x$ as well as $k_y$. This symmetry is unaffected by hexagonal warping. This property also explains the same sign of `legs' occurring for BCP components with increasing warping, as can be observed from columns 1 and 2 of Fig.~\ref{fig:Non_gapped_BPT}. In contrast $G_{xy}$ is odd with respect to $k_x$ and $k_y$.

Nevertheless, Fig.~\ref{fig:Non_gapped_BPT} shows the BCP tensor components in the $k_x-k_y$ plane, with varying warping strengths. As noted, $G_{xx}$ and $G_{yy}$ take arbitrarily large values near the origin, indicated by the white portion of the plots. When warping is zero, we find that $G_{xx}$ and $G_{yy}$ have a double-lobed structure joining near the origin. As warping is turned on and increased, this double-lobed structure distorts and forms `leg' like structures, extending outwards to larger momenta. Increase in warping makes these `legs' more profound, which effectively means the components become less concentrated near the origin and become more spread out. However, the effect of warping is negligible near the origin, which is expected analytically from Equation~\ref{fig:Non_gapped_BPT}, by taking the limit $\lambda\rightarrow0$. Interestingly, $G_{xy}$ takes small values near the origin, a behavior opposite to that of $G_{xx}$ and $G_{yy}$. However, in this case, we find the presence of four `legs' extending from infinity to the origin. In general, we find that the nature of BCP components is substantially different from that of Berry curvature. This suggests that these two quantities manifest in different order phenomena, which turns out to be second and third-order non-linear Hall effect.

The difficulty imposed on obtaining physical insights into the third-order Hall response due to the behavior of BCP tensor components near the origin can be overcome, if a band gap parameter is introduced in our effective model. This is achieved by introducing a parameter $\Delta$ as the coefficient of $\sigma_z$ in the Hamiltonian. The new Hamiltonian for the gapped Rashba system with hexagonal warping now reads

\begin{equation}
\mathscr{H}=\frac{k^2_x + k^2_y + k^2_z}{2m}-(v_y\sigma_xk_y - v_x\sigma_yk_x)+\frac{\lambda}{2}(k^3_+ + k^3_-)\sigma_z+\Delta\sigma_z.
\end{equation}

\begin{figure}[t]
   \centering
    \includegraphics[width=0.9\textwidth]{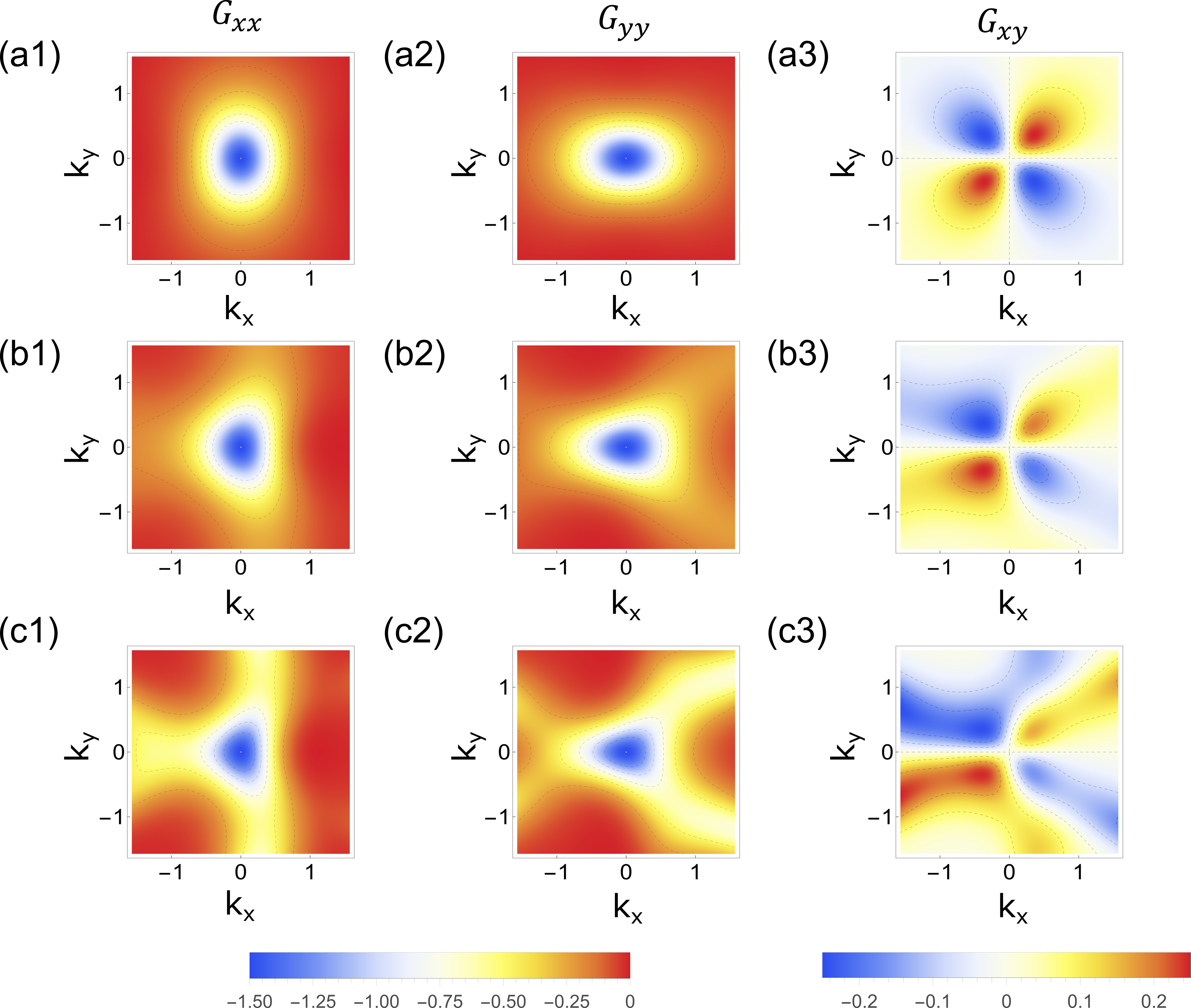}
    \caption{ \textbf{Berry connection polarizability tensor components with gap parameter.} Distribution for Berry connection polarizability tensor components plotted in $k_x-k_y$ plane (with $k_z=0$.) $G_{xx}$ is plotted for (a1) $\lambda=0$, (b1) $\lambda=0.05$, (c1) $\lambda=0.1$; $G_{yy}$ for (a2) $\lambda=0$, (b2) $\lambda=0.05$, (c2) $\lambda=0.1$, and $G_{xy}$ for (a3) $\lambda=0$, (b3) $\lambda=0.05$, (c3) $\lambda=0.09$. For $\lambda = 0$, we recover the usual monopole and quadrupole nature of $G_{xx}$ and $G_{xy}$, respectively. For $\lambda \neq 0$, their nature deviates and new features start appearing. Magnitudes of the components peak where Berry curvature is maximum. Here $\Delta$ is taken to be 0.3 for panels (a1)-(c1) and (a2)-(c2), and $\Delta$=0.25 for panels (a3)-(c3). }
   \label{fig:BPT}
 \end{figure}

Proceeding with the same analysis as before, we calculate the analytical expressions for BCP tensor. As before, the non-zero components of BCP tensor are $G_{xx},G_{yy},G_{xy},$ and $G_{yx}$. The expressions of the BCP tensor components in this case is rather lengthy and cumbersome, therefore we do not write them out explicitly here. However, the plots for the components as shown in Fig.~\ref{fig:BPT} reveal several interesting features. Without the warping, the BCP components exhibit a monopole and quadrupole kind of behavior~\cite{liu2022berry}. With the introduction of warping, however, this monopole or quadrupole like behavior strikingly breaks down. The warping effect is clear from panels (b) and (c) of Fig.~\ref{fig:BPT}. As warping increases, the distribution of BCP components become increasingly asymmetric, and gets elongated along the $k_x$. For the $G_{xy}$ components, we find that for non-zero warping, BCP gets concentrated more in the region with negative $k_x$.

An analytic calculation of the full conductivity tensor involves contribution of several components of $G$ and $\chi$, leading to rather complicated expressions. To gain better insights, we calculate the conductivity tensor after setting $k_z=0$. The third-order Hall response vanishes when electric field is applied along or perpendicular to the mirror line, which in this case is $M_x$. Hence the components $\chi_{yyyx},\chi_{xyyy},\chi_{yxyy},\chi_{yyxy}$ and $\chi_{xxxy},\chi_{xxyx},\chi_{xyxx},\chi_{yxxx}$ will be zero, and will not contribute to the conductivity tensor. So, an in-plane electric field, given by $\mathbf{E}=E(\cos\theta,\sin\theta)$ results in a third-order response $j_{H}^{(3)}=\mathbf{j^{(3)}\cdot \hat{n}}$, where $\mathbf{\hat{n}}$ is the unit vector normal to ${\mathbf{E}}$, given by $(-\sin\theta,\cos\theta)$.

\begin{figure}[t]
    \centering
    \includegraphics[width=\textwidth]{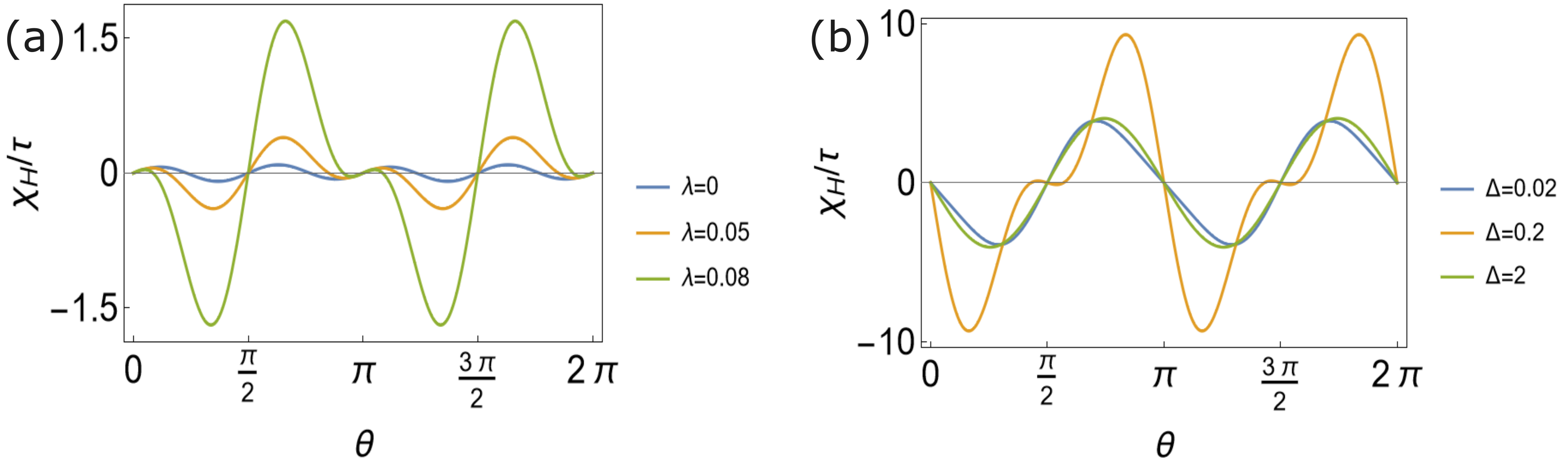}
    \caption{\textbf{Third-order conductivity tensor.} (a) The angular dependence of the third-order conductivity tensor with varying warping strength $\lambda$. In this case $\Delta=0.1$. (b) The angular dependence the third-order conductivity tensor for varying $\Delta$, with $\lambda=0.1$. We observe that as $\lambda$ and $\Delta$ take close values, the maximum magnitude of conductivity increases.}
    \label{fig:chi}
\end{figure}

The third-order transverse conductivity tensor is then given by 

\begin{equation}
    \chi_H(\theta)=\frac{j_{H}^{(3)}}{E^3},
\end{equation}

which can be measured experimentally~\cite{lai2021third}. For systems with mirror symmetry $M_x$, we obtain

\begin{equation}
    \chi_H(\theta)=(-\chi_{11}+3\chi_{21})\sin\theta \cos^3\theta+(\chi_{22}-3\chi_{12})\sin^3\theta\cos\theta,
\end{equation}

with $\theta$ measured from the mirror line, and the shorthand notations being used are $\chi_{11}=\chi_{xxxx}$, $\chi_{22}=\chi_{yyyy}$, $\chi_{21}=(\chi_{yyxx}+\chi_{yxyx}+\chi_{yxxy})/3$, and $\chi_{12}=(\chi_{xxyy}+\chi_{xyxy}+\chi_{xyyx})/3$. We evaluate the third-order conductivity tensor numerically, and it is shown as a function of $\theta$ in Fig.~\ref{fig:chi}. The most remarkable feature of conductivity tensor, in this case, is the interplay of the band gap $\Delta$ and warping term $\lambda$. For all values of $\Delta$ and $\lambda$, we find that $\chi_H$ varies with $\theta$ with a period of $\pi$. Here $\chi_H$ vanishes when $\theta$ is a multiple of $\pi/2$, consistent with our symmetry analysis.

It is worth noting that as the values of $\Delta$ and $\lambda$ become closer, the magnitude of $\chi_H$ increases. Moreover, there is a substantial deviation from the sinusoidal nature, however the periodicity is maintained. This property originates from the structure of $\chi$, and can be used in experimental estimation of warping strengths. To illustrate, suppose we choose a material, in which the band gap can be altered with an external perturbation, such as by the application of an electric field. In such a case, the warping strength is given by the value of band gap where a `resonance' is observed for the maximum value of conductivity. Thus, our proposed hexagonally warped Rashba systems are intriguing platforms for observing and tuning higher-order Hall effects. 

In a very recent independent study~\cite{nag2023third}, results on third-order nonlinear Hall effect are presented, which are complementary to ours. The consistency of results between the two works provides a check about the soundness of the approach used. The major similarities between the two works are the study of BCP tensor components with varying warping strength, and pattern of sign changes between consecutive legs of $G_{xy}$. However, there are important differences between the works worth noting. While Ref.~\cite{nag2023third} uses a tilt term along with the low-energy Hamiltonian, our work introduces a gap parameter, $\Delta$, advantages of which are discussed in the manuscript. Similarly, we study the third order conductivity with the interplay between warping and the band gap. This approach provides intriguing new insights about behaviour of third-order conductivity, and a possible experimental use of the same.

\section{Experimental Consideration}

Based on our calculations, we next estimate the second order nonlinear Hall response that can be observed in experiments. For this, we use the approximation $\omega\tau <<1$, as before. With typical relaxation time in picoseconds, under this approximation, we have

\begin{equation} 
\chi_{xxy}=-\chi_{yxx}=0.185D_{xz}.
\end{equation}

Consider a typical sample size of $10\times5\times1 \;\mu$m$^3$, and isotropic resistivity in all directions to be $50 \;\mu\Omega$cm. Let $V_x=V_y=1$V. Then we have,

\begin{equation}
J^{2\omega}_x=\chi_{xxy}E_xE_y=\chi_{xxy}\frac{V_xV_y}{L_xL_y}= (3.7\times 10^9)D_{xz} 
\; \textrm{A/m}^2.
\end{equation}

The corresponding voltage produced is then $V_x^{2\omega}=0.0185 D_{xz} \;\textrm{V}$. The maximum voltage depending on the warping strength thus is evaluated to be $2.86\times 10^{-4}$ V  for $\lambda=0.1$, $3.77\times 10^{-4}$ V for $\lambda=0.2$, and $11.6\times 10^{-4}$ V for $\lambda=0.3$. These can be measured using standard experimental setups. \\

The third order nonlinear Hall voltage can be estimated by the following relation

\begin{equation}
    V^{(3)}_H\propto\chi_H(\theta)E^3.
 \end{equation}

We consider a typical uniform electric field of 100 V/cm in the $x$ direction. For $\theta=45^o$ and $\Delta=0.1$, keeping other parameters identical to the previous case, the induced voltage magnitude can be estimated to be $V^{(3)}_H\approx 81$ mV for $\lambda=0$, and $V^{(3)}_H\approx 1.22$ V for $\lambda=0.05$ . The interplay between $\Delta$ and $\lambda$ in third-order nonlinear conductivity response can be used in experiments to find the strength of warping in suitable systems. To illustrate, suppose we choose a material, in which the band gap can be altered with an external perturbation, such as by the application of an electric field. In such a case, the warping strength is given by the value of band gap where a ‘resonance’ is observed for the maximum value of conductivity. Thus, our proposed hexagonally warped Rashba systems are intriguing platforms for observing and tuning higher-order Hall effects.

\section{Summary}

In this work, we systematically investigated the Berry curvature multipole physics in Rashba systems with hexagonal warping. We calculated the general expressions for Berry curvature, using a low-energy model, and studied how the model parameters affect its distribution. We next investigated the behaviour of Berry curvature dipole in these systems, and found an unconventional dipole profile. This could be explained using the band structure and the Berry curvature distribution. Additionally, we looked at the dipole density profiles and concluded that the warping of the band structure itself is an important factor in determining the dipole profile. Motivated by the low-energy model results, we used a tight-binding model to study the band structure and the BCD in a more realistic setting. Here, we found another important factor for the unusual dipole profile, which is the particle-hole asymmetry. Further, we performed analytic calculations for Berry connection polarizability tensor, which is related to the third-order Hall effect. We found that for our Rashba systems with hexagonal warping, four of the nine components are non-zero. Further, by introducing a band gap in our model, we found interesting connections of the band gap with the warping strength, and their combined effect on the third-order Hall response. We hope that our work can help in the experimental realization of Berry curvature multipole physics in Rashba materials with hexagonal warping, and provide a new avenue for engineering non-linear Hall effects.

In an independent work, the second-order nonlinear Hall effect was studied using a low-energy
model~\cite{yar2022nonlinear}. Our work goes beyond this by the use of a tight-binding
model in addition to the low-energy Hamiltonian. Tight binding model provides much more realistic insights into the material properties, and brings out crucial physics that is often not seen in
  the use of low-energy model~\cite{zeng2021nonlinear}. In this independent work~\cite{yar2022nonlinear}, the BCD density is studied using a low-energy model similar to ours. However, there are crucial differences between our work and Ref.~\cite{yar2022nonlinear}. In the context of BCD density, our work studies the relationship between BCD density and warping strength. This is in contrast to~\cite{yar2022nonlinear}, which focuses on a single warping strength. Considering different warping strengths also provides intriguing connections to the way BCD is distributed near the origin ($k_x\approx0$ and $k_y\approx0$).

\section*{Acknowledgments}
We acknowledge useful discussions with S. Roy, D. Varghese, A. Bandyopadhyay, A. Bose, and A. Banerjee. S. S. thanks the Kishore Vaigyanik Protsahan Yojana (KVPY) for a fellowship. A. N. acknowledges the startup grant of the Indian Institute of Science (SG/MHRD-19-0001).

\section*{Appendix}

The eigenvectors corresponding to the eigenvalues of Eqn. (14) are given, upto normalisation constant and a phase, by

\begin{equation}
   \psi_{\pm}=\begin{pmatrix} k_x^3 \lambda -3k_x k_y^2\lambda\pm\sqrt{k_x^2 v_x^2+k_y^2 v_y^2 +k_x^6\lambda^2-6k_x^4k_y^2\lambda^2+9k_x^2k_y^4\lambda^2} \\ -k_y v_y + i k_x v_x \end{pmatrix}.
\end{equation}

The group velocity terms, $\frac{\partial H}{\partial k}$, are given by

\begin{subequations} 

   \begin{equation}
    \frac{\partial H}{\partial k_x}=  \begin{pmatrix}
         k_x/m+3\lambda(k_x^2-k_y^2)& -iv_x \\
         iv_x & k_x/m-3\lambda(k_x^2-k_y^2)
    \end{pmatrix},
    \end{equation}

   \begin{equation}
     \frac{\partial H}{\partial k_y}= \begin{pmatrix}
         k_y/m-6k_yk_x\lambda& -v_y \\
        -v_y & k_y/m+6k_yk_x\lambda 
    \end{pmatrix},
   \end{equation}
   
    \begin{equation}
      \frac{\partial H}{\partial k_z}= \begin{pmatrix}
             k_z/m & 0 \\
            0 & k_z/m 
        \end{pmatrix}.
     \end{equation}

     \end{subequations}
    
Plugging the eigenvectors and the group velocity terms in Eqn. 4, the expressions for Berry curvature follow.\\

In order to derive the tight-binding Hamiltonian (Eqn. 18), we use the approximations $\sin{k_x}\approx k_x$, $\cos{k_x}=1-\frac{k_x^2}{2}$, and similar for $k_y$ and $k_z$. Thus, the various terms of the low-energy Hamiltonian transform, as follows

\begin{subequations}
\begin{equation}
  \frac{1}{2m} (k_x^2+k_y^2+k_z^2) = \frac{1}{2m} (2-2\cos{k_x}+2-2\cos{k_y}+2-2\cos{k_z})= \frac{1}{2m} (6-2\cos{k_x}-2\cos{k_y}-2\cos{k_z}),            
\end{equation}

\begin{equation}
    -v_yk_y\sigma_x = -v_y\sin{k_y}\sigma_x,
\end{equation}

\begin{equation}
    v_xk_x\sigma_y = v_x\sin{k_x}\sigma_y,
\end{equation}

\begin{equation}
 \begin{split}
    \frac{\lambda}{2}(k_+^3+k_-^3)\sigma_z & =   \frac{\lambda}{2}\times2k_x(k_x^2-3k_y^2)\sigma_z = \lambda\sin{k_x}((2(1-\cos{k_x})-3(2(1-\cos{k_y}))))\sigma_z \\
   & = \lambda\sin{k_x}(6\cos{k_y}-2\cos{k_x}-4)\sigma_z.
    \end{split}
\end{equation}
\end{subequations}

\bibliography{Ref.bib}

\end{document}